\documentclass[aps,prd,twocolumn,showpacs,superscriptaddress,nofootinbib,preprintnumbers]{revtex4-2}

\usepackage[pdftex,colorlinks]{hyperref}
\usepackage[dvipsnames]{xcolor}
\definecolor{phthaloblue}{rgb}{0.0, 0.06, 0.54}

\hypersetup{
    colorlinks=true,
    linkcolor=MidnightBlue,
    citecolor=MidnightBlue,
    filecolor=MidnightBlue,
    urlcolor=MidnightBlue,
    }
\usepackage[pdftex]{graphicx}
\usepackage{bm}
\usepackage{cancel}
\usepackage{here}
\usepackage{comment,braket}
\usepackage{epsf}
\usepackage{amsmath}
\usepackage{graphics}
\usepackage{amsfonts}
\usepackage{amssymb}
\usepackage{latexsym}
\usepackage{color}
\usepackage{natbib}
\usepackage[normalem]{ulem}
\input{colordvi.tex}
\usepackage{feynmp}
\usepackage[utf8]{inputenc}
\DeclareUnicodeCharacter{2212}{-}
\usepackage{orcidlink}
\newcommand{\Ap}{A^\prime}

 \newcommand{\brac}[2]{ \left( \frac{#1}{#2} \right) } 

\newcommand{\be}{\begin{eqnarray}}
\newcommand{\ee}{\end{eqnarray}}

\newcommand{\beq}{\begin{equation}}
\newcommand{\eeq}{\end{equation}}

\newcommand{\Eq}[1]{Eq.~(\ref{#1})}

\begin{document}

\preprint{\tt FERMILAB-PUB-24-0554-T}

\title{Discovering Dark Matter with the MUonE Experiment}

\author{Gordan Krnjaic\,\orcidlink{0000-0001-7420-9577}}
\email{krnjaicg@fnal.gov}
\affiliation{Theoretical Physics Division, Fermi National Accelerator Laboratory, Batavia, IL, USA}
\affiliation{Department of Astronomy \& Astrophysics, University of Chicago, Chicago, IL USA}
\affiliation{Kavli Institute for Cosmological Physics, University of Chicago, Chicago, IL USA}

\author{Duncan Rocha\,
\orcidlink{0000-0002-8263-7982}}
\email{drocha@uchicago.edu}
\affiliation{Theoretical Physics Division, Fermi National Accelerator Laboratory, Batavia, IL, USA}
\affiliation{Department of Physics, University of Chicago, Chicago, IL USA}

\author{Isaac R. Wang\,\orcidlink{0000-0003-0789-218X}}
\email{isaacw@fnal.gov}
\affiliation{Theoretical Physics Division, Fermi National Accelerator Laboratory, Batavia, IL, USA}

\date{\today}

\begin{abstract}
    The MUonE experiment aims to extract the hadronic contribution to
    the muon anomalous magnetic moment from a precise measurement of
    the muon-electron differential scattering cross section. 
    We show that MUonE can also discover thermal relic dark matter using only its nominal experimental setup. Our search strategy is sensitive to models of dark matter in which pairs of pseudo-Dirac fermions are produced in muon-nucleus scattering in the target, and the heavier state decays semi-visibly to yield dilepton pairs displaced downstream from the interaction point. This approach can probe sub-GeV thermal-relic dark matter whose cosmological abundance is governed by the same model parameters that set the MUonE signal strength. Furthermore, our results show that the downstream ECAL plays a key role in rejecting backgrounds for this search, thereby providing strong motivation for the MUonE to keep this component in the final experimental design.
\end{abstract}

\bigskip
\maketitle

{\bf Introduction.}
There has recently been great interest in developing new search strategies for dark matter discovery at fixed target accelerators involving proton \cite{deNiverville:2016rqh,Izaguirre:2015pva}, electron \cite{LDMX:2018cma}, and muon \cite{NA64:2024klw} beams -- see Refs. \cite{Gori:2022vri,Battaglieri:2017aum} for detailed summaries. 
Collectively, these efforts aim to probe models in which sub-GeV DM achieves a thermal relic abundance through direct annihilation to Standard Model (SM) final states, such that the signal strength in the laboratory is governed by the same parameters that set the DM cosmological abundance.

The proposed MUonE experiment at CERN~\cite{Abbiendi:2016xup,Abbiendi:2677471} aims to measure the angular distribution of elastic muon-electron scattering to extract the hadronic contribution to the muon anomalous magnetic moment \cite{Muong-2:2006rrc}.
In this setup, a 160 GeV muon beam impinges on a series of thin, beryllium target modules surrounded by layers of tracking material. When muons upscatter stationary electrons in these targets, the angular trajectories of final state particles can be resolved within $\sim 0.02$ mrad.
This capability also makes MUonE an excellent probe of new forces in muon-electron \cite{Galon:2022xcl,Asai:2021wzx} or muon-nucleon \cite{GrillidiCortona:2022kbq} scattering, provided that the force carriers decay to yield displaced vertices sufficiently far from the beryllium target layers.

\begin{figure}[t!]
    \includegraphics[width=1.0\linewidth]{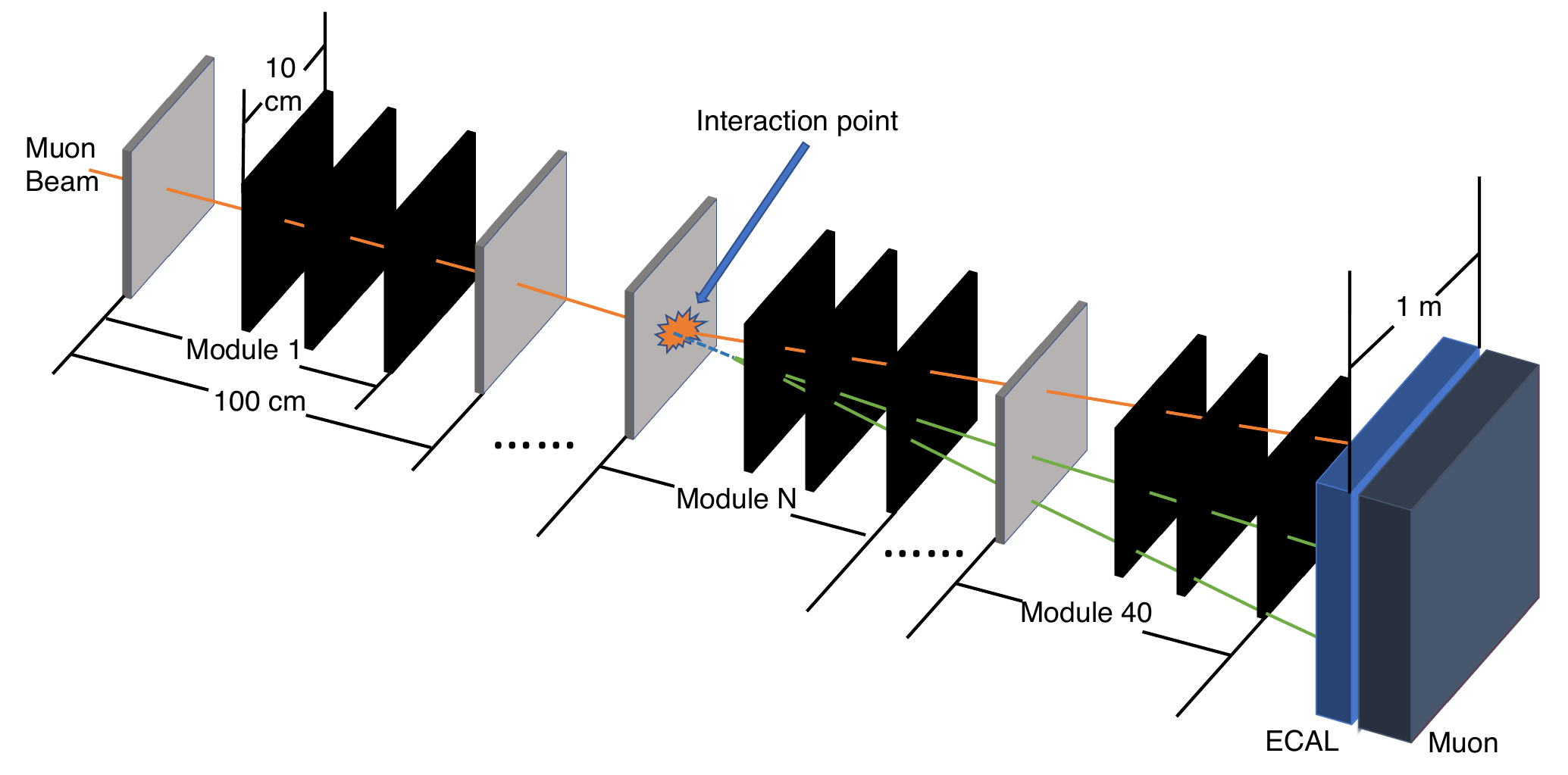}
    \caption{Schematic diagram of inelastic dark matter production in muon-nucleus scattering at the MUonE experiment. An incoming 160 GeV muon beam (orange line) scatters a beryllium nucleus at the interaction point and produces pseudo-Dirac dark states $\chi_1$ and  $\chi_2$ through the processes depicted in Fig. \ref{fig:cartoon}.  The heavier of these states decays semi-visibly via $\chi_2 \to \chi_1 \ell^+\ell^-$ such that the dileptons  (green lines) emerge from a displaced vertex. Here the gray sheets represent the target material and the black sheets are tracking layers. The downstream ECAL serves to reject SM background processes. Image adapted from Ref.~\cite{Galon:2022xcl} and modified for our signal process with a different final state.}
    \label{fig:scheme}
\end{figure}

In this {\it Letter}, we show for the first time that MUonE can be sensitive to thermal relic dark matter without any modifications to the proposed experimental design.
Our benchmark scenario is a model of inelastic DM in which a pseudo-Dirac fermion pair with unequal masses couple to a kinetically mixed dark photon~\cite{Tucker-Smith:2001myb}. The heavier of these states is unstable and decays semi-visibly to yield displaced tracks of charged particles on laboratory length scales. At MUonE, both states are radiatively produced in muon-nucleus interactions inside the target layers and the heavier state decays in the forward direction to yield missing energy and dilepton pairs downstream in the tracking layers as depicted schematically in Fig. \ref{fig:scheme}.

\medskip 

\begin{figure*}[t!]
    \centering
    \hspace{-0.5cm}
    \includegraphics[width=0.8\linewidth]{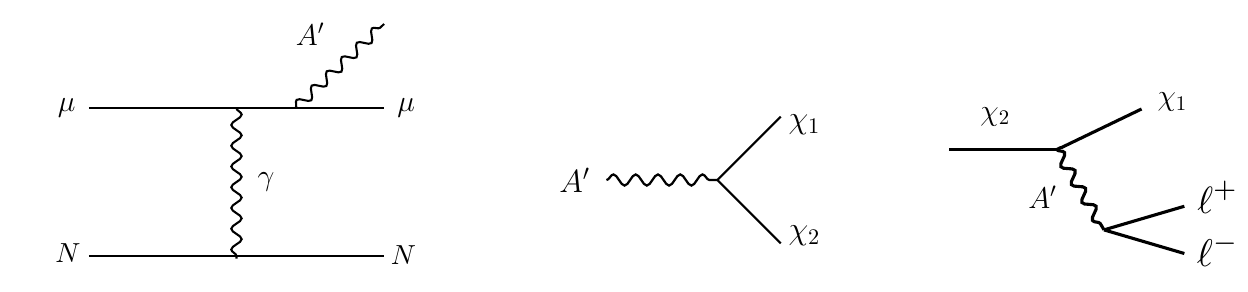}
    \caption{Feynman diagrams representing the sequence of steps that yield the inelastic DM signal at MUonE. {\bf Left:} a dark photon $A^\prime$ is radiatively produced in $\mu$-$N$ scattering, where $N$ is a Be nucleus. {\bf Middle:} the $\Ap$ decays through the off-diagonal coupling to the pseudo-Dirac fermions. {\bf Right:} for $\Delta < m_{\Ap},$ the heavier state decays via $\chi_2 \to \chi_1 \ell^+\ell^-$ to yield a displaced dilepton pair downstream of the interaction point at MUonE.}
    \label{fig:cartoon}
\end{figure*}

{\bf Inelastic Dark Matter.}
If a dark matter fermion has a dominant Dirac mass and a small Majorana mass, diagonalizing its mass matrix yields two distinct eigenstates with a mass splitting $\Delta$ \cite{Tucker-Smith:2001myb}. If the original fermion couples to the Standard Model (SM) through a vector current, in the mass basis this interaction becomes off-diagonal with several key implications: 
\begin{itemize}
\item{\bf Unstable Heavier State:} The off-diagonal coupling through the vector current mediates decay reactions for the heavier state. Thus, the dark matter today consists entirely of the stable lighter state. 
    \item {\bf Inelastic Scattering:} In order for the lighter state to scatter off SM targets, it must undergo an inelastic transition. Since DM is non-relativistic in our galaxy, for sufficiently large $\Delta$, upscattering into the heavier state becomes kinematically forbidden and there are no direct detection signals.   

       \item {\bf Co-annihilation:} Since the only dark sector interaction is off-diagonal, the two states cannot annihilate themselves, but can {\it co-annihilate} each other. Since the heavier state is generically absent in the present-day universe, there are no indirect-detection prospects for this scenario.\footnote{However, for small values of $\Delta$, the heavier state can be revived in the galaxy, such that coannihilation can occur \cite{Berlin:2023qco}} Furthermore, coannihilation is fully compatible with the predictive thermal freeze-out mechanism and can extend its viability  below masses of ${\cal O}(10 \, \rm GeV)$ \cite{CarrilloGonzalez:2021lxm} despite otherwise strong bounds on from CMB energy injection bounds at low mass \cite{Slatyer:2009yq,Planck:2018vyg}.

 \item {\bf Displaced Vertices at Accelerators:} Although direct and indirect detection are generically unavailable for this scenario, accelerator production remains a viable and promising discovery strategy since both light and heavy states can be produced together with relativistic kinematics  \cite{Izaguirre:2015zva,Izaguirre:2017bqb,Jordan:2018gcd,Berlin:2018jbm,Duerr:2020muu}. Since the heavier
 state is unstable, it may decay on accelerator length scales to yield distinctive displaced vertices. Generically, the same couplings that govern the lifetime of the heavier state also set the thermal relic abundance in these models.
\end{itemize}

\noindent Thus, since other experimental probes are 
generically unavailable for this class of models, accelerator production may be the only way to test thermal freeze-out within this framework. In the remainder of this {\it Letter}, we explore inelastic production DM at MUonE within the context of a benchmark model, but we emphasize that our strategy can apply to a variety of scenarios with these features.

{\bf Representative Model.} For our benchmark scenario, we couple a pseudo-Dirac fermion pair $\chi_1$ and $\chi_2$ to a kinetically mixed dark photon $\Ap$ through an off-diagonal interaction 
\be
\label{eq:main_lag}
{\cal L}_{\rm int} = - \epsilon e A_\mu^\prime J^{\mu}_{\rm EM} -  (g_D A_\mu^\prime \overline \chi_2 \gamma^\mu \chi_1 + {h.c.} )~~,
\ee
where $A^\prime$ is a massive ``dark photon" mediator with mass $m_{A^\prime}$ and kinetic mixing parameter $\epsilon$, $J_{\rm EM}$ is the SM electromagnetic current, and the $\chi_{1,2}$ have masses $m_{1,2}$ with splitting $\Delta \equiv m_2 - m_1$. The off-diagonal coupling  $\Ap \chi_2 \gamma^\mu \chi_1$ in \Eq{eq:main_lag} can naturally arise if a four-component fermion has a Dirac mass and a Majorana mass in the interaction basis, such that its vector current with $\Ap$ is off-diagonal in the mass eigenbasis $\chi_{1,2}$\cite{Tucker-Smith:2001myb}.

In the predictive parameter space of interest,\footnote{in the opposi'te regime where $\epsilon e \gtrsim g_D$, achieving the observed relic abundance requires large values of $\epsilon$ that are largely excluded by existing experiments. Furthermore, if $m_{\Ap} < m_1 + m_2$, the relic abundance arises from $\chi_i \chi_i \to \Ap \Ap$ annihilation, and does not depend on $\epsilon$ which governs accelerator production. }  $g_D\gg e \epsilon$ and $m_{\Ap} > m_1 + m_2$, so the dark branching fraction satisfies Br$(\Ap \to \chi_1 \chi_2) \approx 1$.  If $\Delta < m_{\Ap}$, the unstable heavier state decays via $\chi_2 \to \chi_1 \bar f f$ where, for each $f$ channel, which corresponds to the decay length\footnote{Here we have used the fact that the $\chi_2$ rest frame lifetime satisfies $\tau =  15\pi m_{\Ap}^4/(4 \epsilon^2 \alpha \alpha_D \Delta^5)$ \cite{CarrilloGonzalez:2021lxm}}
\be
\label{eq:lifetime}
 \frac{c \tau}{\gamma}
  \approx  10 \, {\rm cm} 
\brac{m_{\Ap}}{10^2 \rm \, MeV}^{ \! 4} \!
\brac{20 \rm \, MeV}{\Delta}^{ \! 5} \! \!
\brac{10^{-7}}{\epsilon^2 \alpha_D},~~ \nonumber
\ee
where $\gamma$ is the boost factor and we have taken the massless $f$ limit. In principle $f$ here could be any fermion, but for the $\Delta$ of interest here, most of the MUonE signal
arises from $\chi_2 \to \chi_1 \ell^+\ell^-$ decays, where $\ell = e, \mu$.

In the early universe at temperatures $T \gg m_{1,2}$, the $\chi_{1,2}$ maintain chemical equilibrium with the SM through $\chi_1 \chi_2 \leftrightarrow \bar f f$ coannihilation through virtual $\Ap$ exchange, where $f$ is a charged SM fermion. In the  $m_{\Ap} \gg m_1 \gg \Delta, m_f$ limit, the annihilation cross section times velocity for a single $f$ final state scales as \cite{Izaguirre:2015pva}
\be
\label{eq:sigmav}
\sigma v \propto 
\frac{\epsilon^2 \alpha_D m_1^2}{m_{\Ap}^4} 
\equiv \frac{y}{m_1^2}~,~ y \equiv \epsilon^2 \alpha_D \brac{m_1}{m_{\Ap}}^4,
\ee
where $y$ is a dimensionless interaction strength parameter and, in the same limit, the relic density satisfies \cite{Berlin:2018bsc}
\be
\Omega_{\chi_1} + \Omega_{\chi_2} \sim 0.1 \brac{y}{10^{-10}} \brac{100 \, \rm MeV}{m_1}^2~.
\ee
For the larger (order unity) mass splittings we consider here, the relic density is obtained by numerically solving the Boltzmann equations for this system, as done in Refs. \cite{Izaguirre:2015pva,CarrilloGonzalez:2021lxm,Berlin:2018jbm,Jordan:2018gcd} and we use these curves from Ref. \cite{Berlin:2018jbm} for comparison with the MUonE sensitivity.

\begin{figure}[t!]
    \centering
    \includegraphics[width=0.85\linewidth]{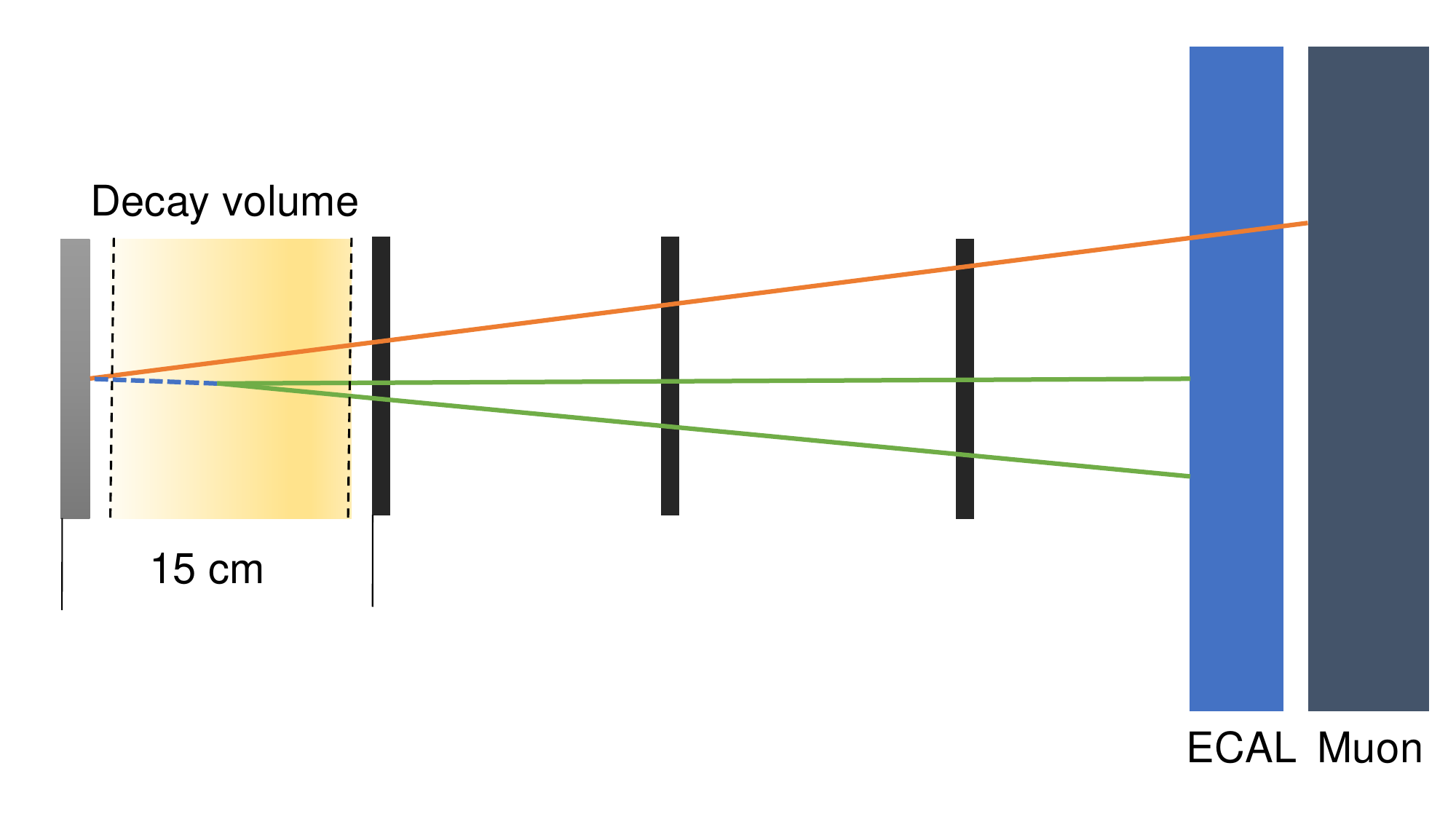}
    \caption{A schematic depiction of the inelastic DM signal in one MUonE target module. Here the red track is the beam muon and the green tracks which reconstruct a displaced vertex are the additional charged tracks from the displaced $\chi_2 \to \chi_1 \ell^+\ell^-$ decay.  We require that the $\chi_2$ decay occur within the $\sim 15$ cm fiducial decay volume (shaded yellow), whose longitudinal endpoints satisfy the inequalities in \Eq{eq:deltaz}. Image adapted from Ref. ~\cite{Galon:2022xcl} and modified to represent our signal process. }
    \label{fig:close}
\end{figure}

After $\chi_{1,2}$ chemically decouple from the SM,
the heavier $\chi_2$ state is further depleted by decays as long as $\Delta > 2m_e$, which is the relevant parameter space in this work. Thus, for $T\ll \Delta$, the late time DM population consists  only of $\chi_1$ particles and no $\chi_2$ coannihiltion partners. Thus, during the CMB epoch, this model evades limits on late-time energy injection, which otherwise rule out thermal relics with mass below 20 GeV \cite{Planck:2018vyg}.  In the present-day universe, $\chi_1$ constitutes all of the cosmological and galactic dark matter. 
For a detailed discussion of this model, its cosmological history, and various experimental constraints, see  Refs. \cite{Izaguirre:2015zva,Izaguirre:2017bqb,Berlin:2018bsc,Berlin:2018jbm,CarrilloGonzalez:2021lxm}

In the remainder of this {\it Letter}, we describe our MUonE search strategy in the context of the specific model presented here. However, the generic features of our approach naturally generalize to other scenarios with different mediators and final state particles (see Ref. \cite{Bauer:2017qwy} for a discussion of other mediators and their limits). We leave these analyses for future work.

\begin{figure*}[t!]
\includegraphics[width=0.4\linewidth]{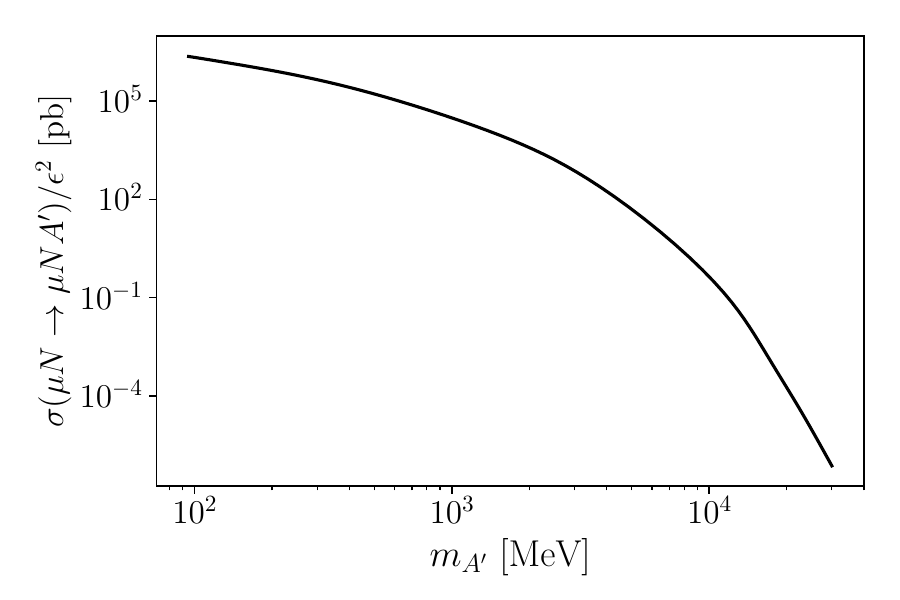}~~
\includegraphics[width=0.4\linewidth]{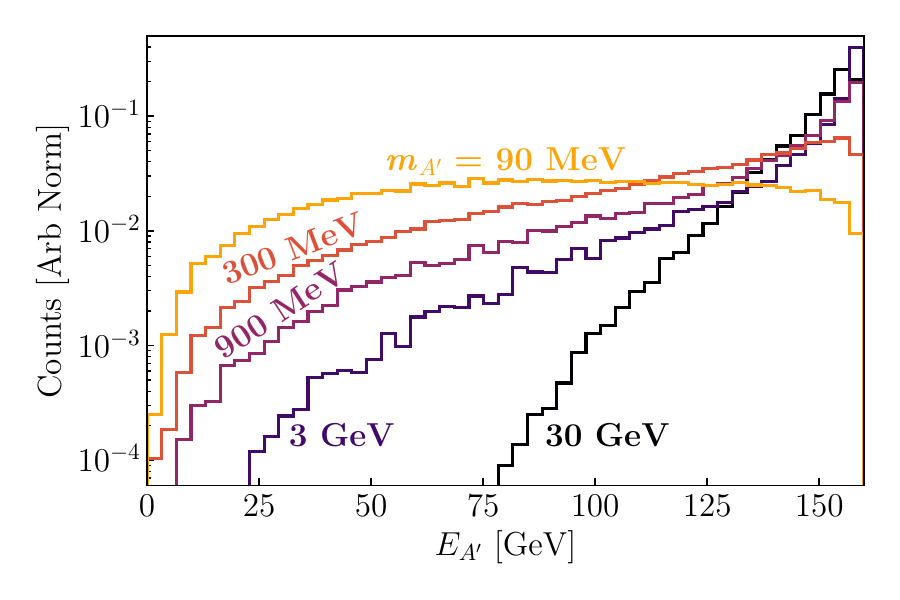}
    \caption{ {\bf Left: }
    Cross section for dark photon production via the bremmstrahlung-like process $\mu N \to \mu N A'$ at MUonE, where $N$ is a beryllium nucleus, calculated using \texttt{CalcHEP}. 
    {\bf Right:} 
    Recoiling $\Ap$ energy distribution 
    from the same simulation sample prior to any event selection.
   }
    \label{fig:aprime-dist}
\end{figure*}

\begin{figure*}
\includegraphics[width=0.335\linewidth]{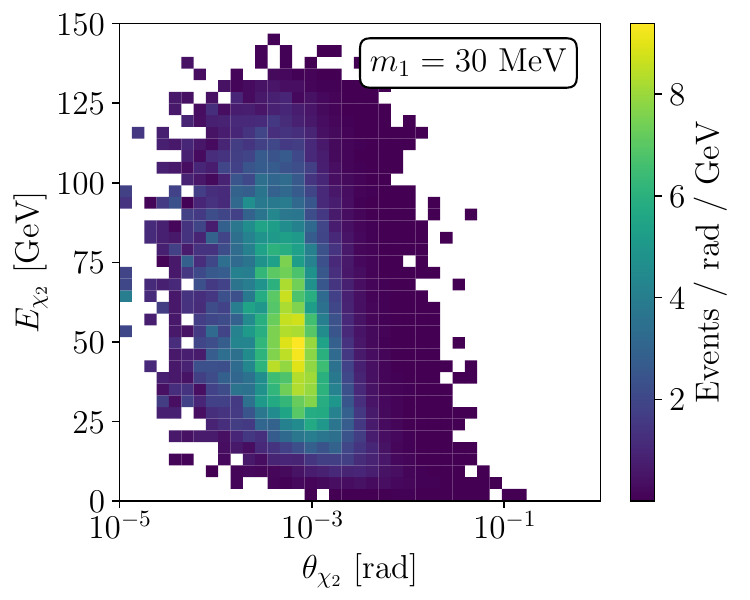}~~
\includegraphics[width=0.335\linewidth]{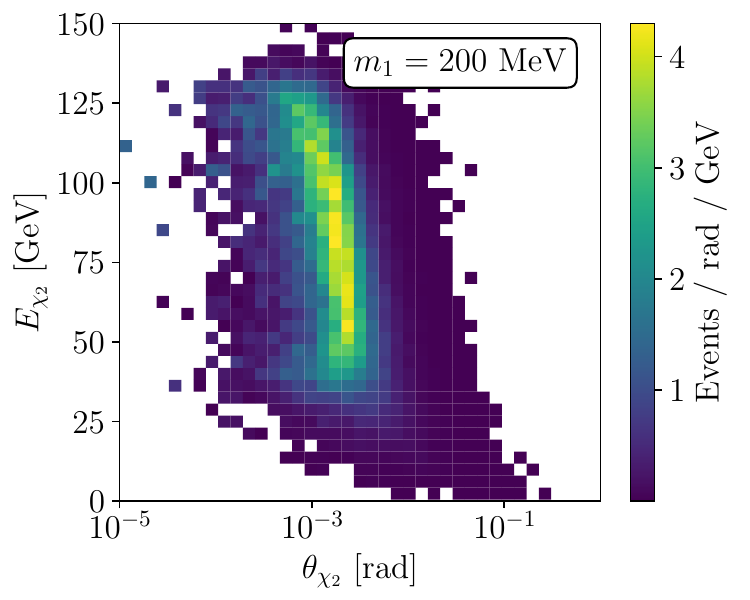}~~
\includegraphics[width=0.335\linewidth]{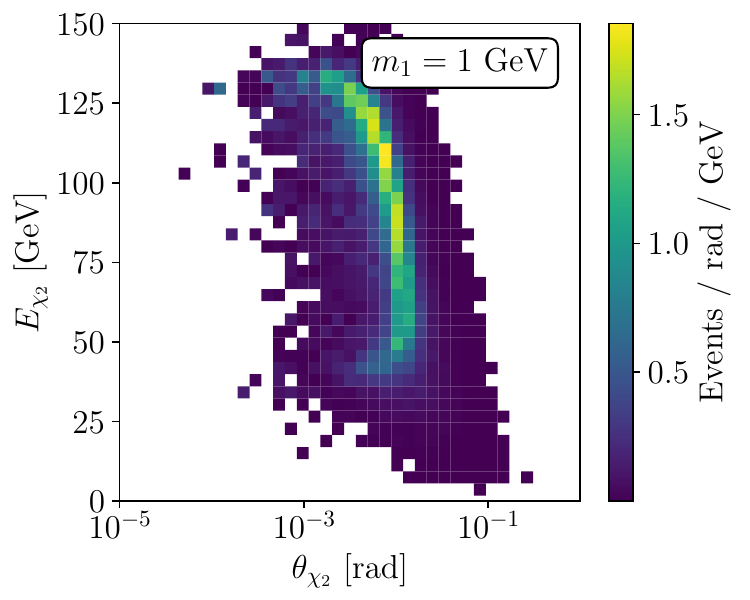}
    \caption{ Energy angle distributions for $\chi_2$  
    from our simulation of $\mu N \to \mu N \Ap$ scattering, followed by $\Ap \to \chi_1\chi_2$ decays at MUonE for various values of $m_1$ assuming $\Delta = 0.4 m_1$, prior to any event selection.
      }
    \label{fig:x2-dist}
\end{figure*}

{\bf The $\text{MUonE}$ Experiment.} 
Inspired by NA7 experiment~\cite{Amendolia:1984nz,NA7:1986vav},  
MUonE aims to extract the running $\alpha$ by precisely measuring the differential 
    $\mu^\pm e^- \to \mu^\pm e^-$
cross section ~\cite{Abbiendi:2016xup,Abbiendi:2677471}.
This running can be used to extract the hadronic vacuum polarization contribution to muon anomalous magnetic magnetic moment ~\cite{Borsanyi:2020mff,CMD-3:2023alj}, which is crucial to resolving the longstanding discrepancy between theory \cite{Aoyama:2020ynm,Boccaletti:2024guq} and experiment~\cite{Muong-2:2006rrc,Muong-2:2021ojo,Muong-2:2023cdq}.
For an up-to-date status report on MUonE, see Ref.~\cite{Spedicato:2024qjd}.

MUonE plans to deliver a 160 GeV muon beam from the CERN M2 beamline onto a series of 40 target modules. Each module consists of one 15 mm Be sheet and three downstream tracking layers with $10 \times 10~\rm cm^2$ cross-sectional area. Within each module, the first tracking layer is positioned 15 cm behind the target. The exact location of the 2nd and 3rd tracking layers has not been fully decided yet.
In this work, we apply 1 meter as the distance from the target to the 3rd tracking layer as a conservative selection criterion for angular acceptance and assume the 2nd tracking layer stands at the middle point between the 1st one and the 3rd one.
The target modules are spaced 1 m apart from each other along the beamline and this setup is expected to achieve an integrated luminosity of
     $\mathcal{L} = 1.5 \times 10^4~\rm pb^{-1}$
for $\mu$-$e$ scattering, corresponding to $\sim 10^{16}$ muons on target.

\begin{table}[t!]
    % \centering
    \begin{tabular}{cc}
         \hline
        Variable                     & Selection Criteria        \\
        \hline
        Decay $z$ coordinate           & $25$\,mm  $< z < 140$\,mm
        \\
        Decay daughter energy     & $> 5$\,GeV
        \\
        Decay daughter opening angle & $> 1$ mrad \\
        Charged track geometry & Hit all 3 trackers \\
        Modules & Last 5 \\
        \hline 
    \end{tabular}
    \caption{Event selection criteria for our proposed search. In our numerical results, these requirements are imposed on our signal and background MC events. These criteria ensure that SM backgrounds are negligible for our signal of interest (see Supplemental Material for a discussion).}
    \label{tab:cuts}
\end{table}

To achieve high measurement accuracy, MUonE employs a CMS-based tracking apparatus~\cite{Abbiendi:2677471,CERN-LHCC-2017-009,Migliore:2797715}, which can resolve the angle resolution of outgoing particles to within $0.02~\mathrm{mrad}$.
There is no applied magnetic field, so the trajectories of outgoing charged particles are not bent as they traverse the beamline. Note that the experimental setup is not designed to measure the outgoing energies or momenta of any charged particles passing through the instrumented region. 

Behind the last target module, there is a proposed electromagnetic calorimeter (ECAL), followed by a muon filter system.
The precise particle identification (PID) efficiency and overall size of this combined system have not yet been finalized, but its cross-sectional area is expected to be of order $1 \times 1 ~\rm m^2$~\cite{Abbiendi:2677471};
for concreteness, here we assume this area is $1 \times 1~\rm m^2$.
In Fig.~\ref{fig:scheme}, we show a schematic design of MUonE experiment with particle trajectories representing our inelastic DM signal process.\footnote{ Unlike the cartoon in Fig. \ref{fig:scheme}, the nominal MUonE signal of interest would consist of only one muon track and one electron track downstream of the interaction point}

{\bf  Dark Matter Signal.}
In our scenario, the MUonE signal arises from the steps depicted in Fig. \ref{fig:cartoon} and corresponds to the sequence
\be
\label{eq;signal-topology}
\mu N  & \to \mu N A' ~~,    ~~
A'     & \to \chi_1 \chi_2~~,~~\chi_2  \to \chi_1 \ell^+ \ell^-,~~
\ee
where $N$ is a beryllium nucleus in the target. Since the $\chi_1$ is invisible, the downstream signal arises purely from $\chi_2$ decays,
which generically yield a displaced vertex in our parameter space of interest. Since $c\tau \propto m_{\Ap}^4/(\Delta^5 \epsilon^2 \alpha_D)$ from \Eq{eq:lifetime}, even mild hierarchies in the $m_{\Ap}/\Delta > 1$ ratio can yield macroscopic displacements. Note that the same model parameters that govern
the dark matter relic density
also set the signal strength at MUonE, and therefore provide an experimental milestone for discovering or falsifying this scenario in the laboratory.

\begin{figure*}
        \includegraphics[width=0.33\textwidth]{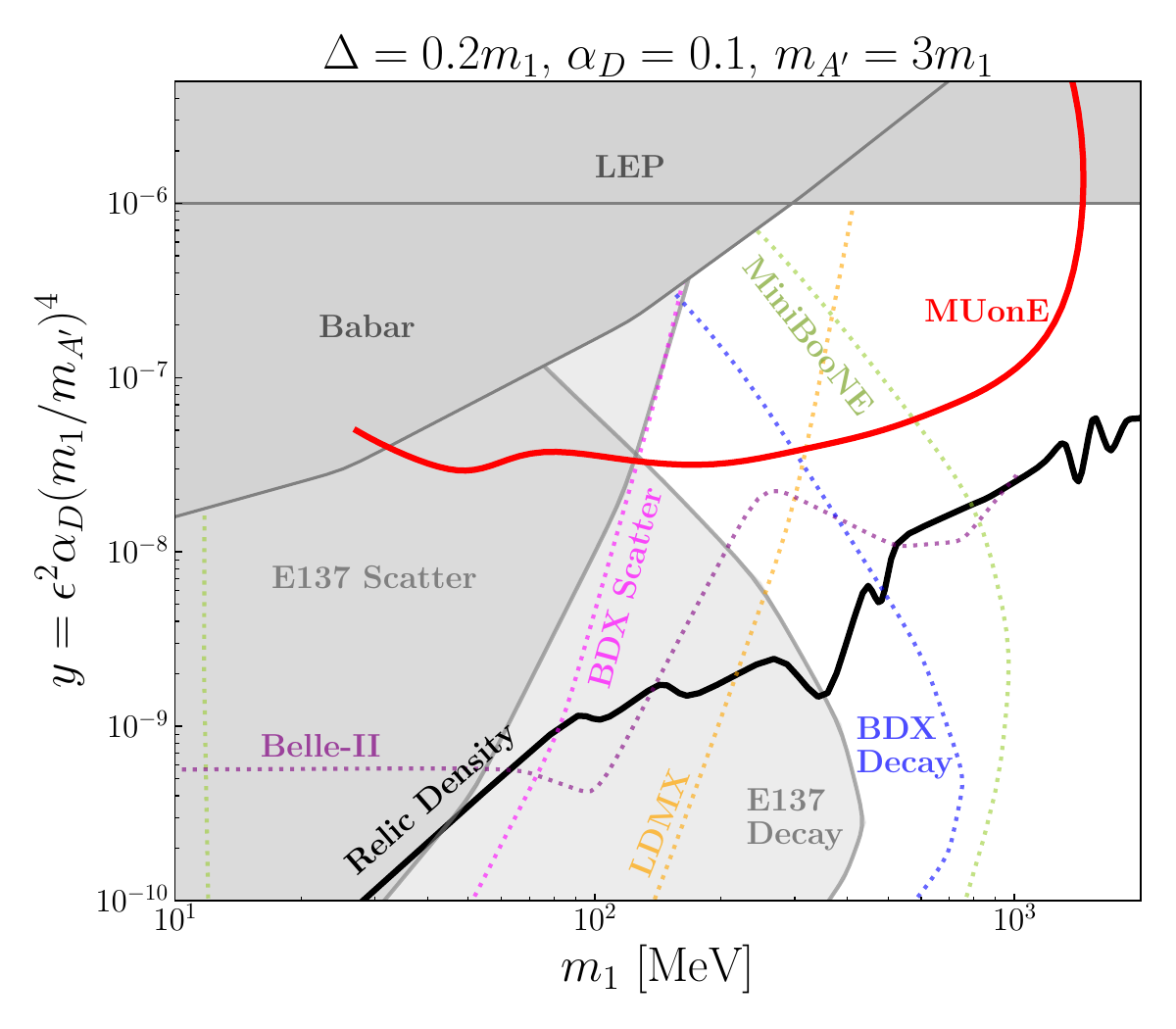}~
        \includegraphics[width=0.33\textwidth]{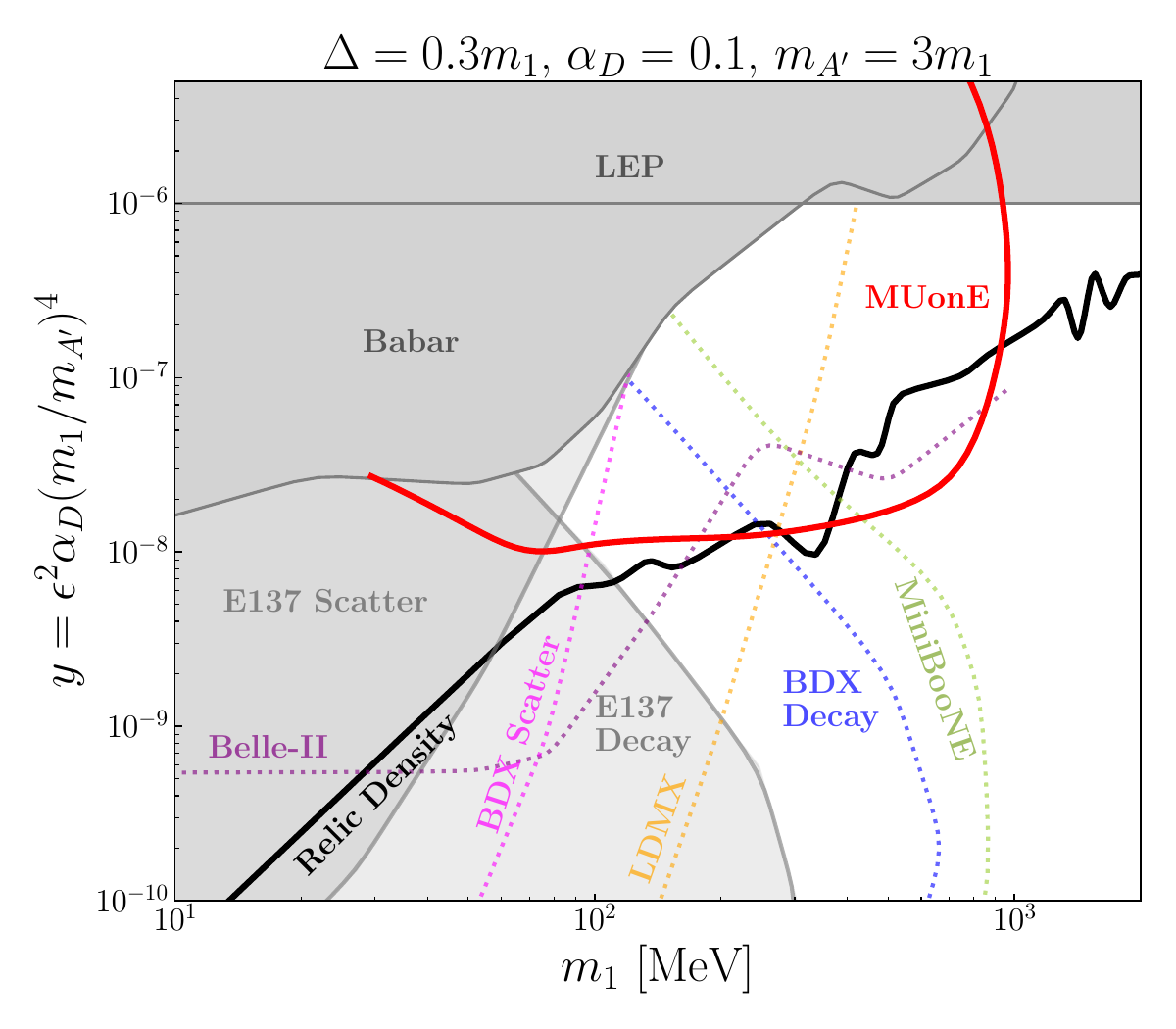}~~
        \includegraphics[width=0.33\textwidth]{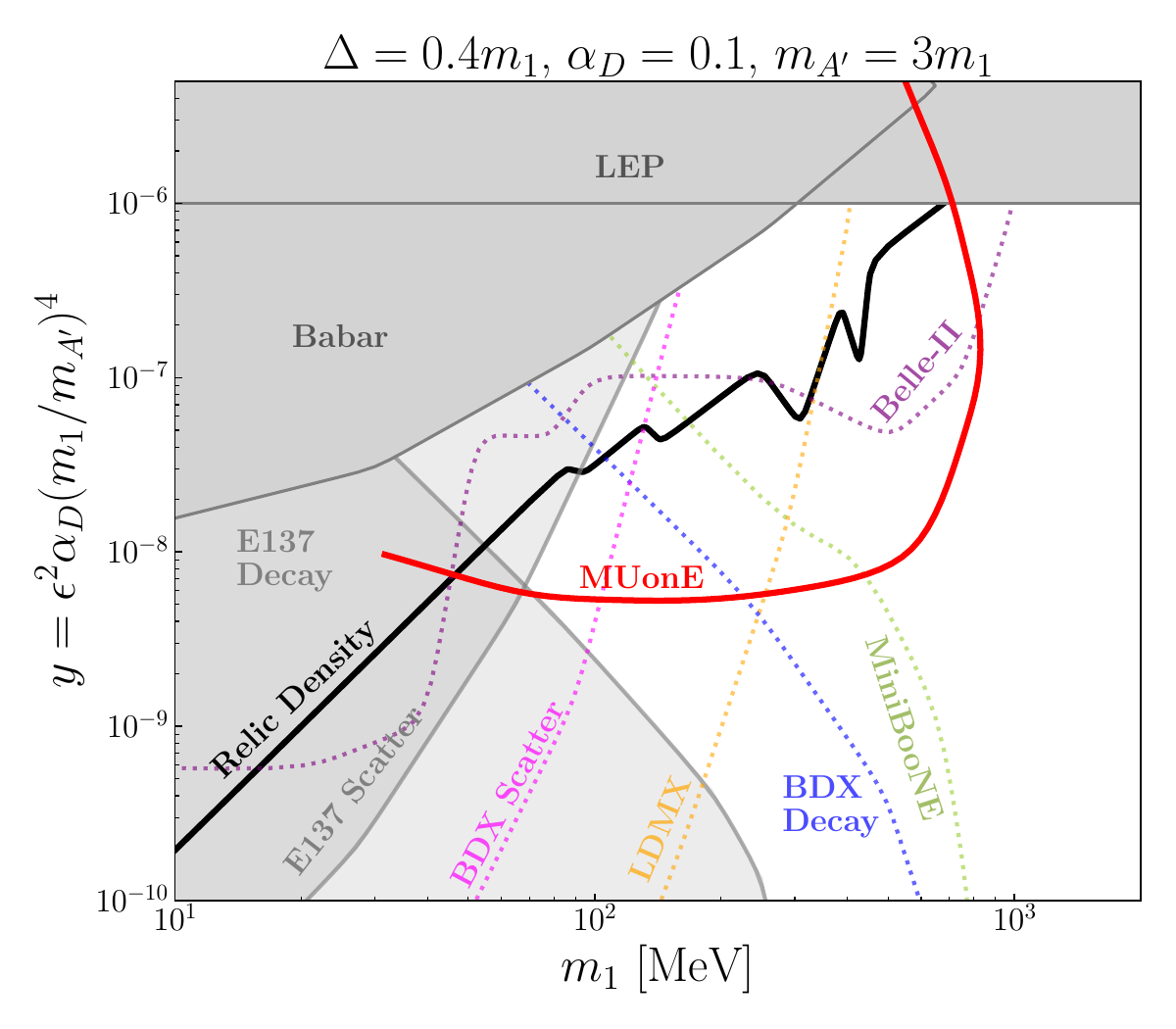}
    \caption{Projected inelastic DM sensitivity at the MUonE experiment (red curve) for a variety of mass splittings assuming a muon luminosity of 
    $4.7 \times 10^3$ pb$^{-1}$. Along the black curve, the DM achieves the observed relic density through $\chi_1 \chi_2$ coannihilation \cite{Berlin:2018bsc}.  The gray-shaded regions are excluded by BABAR \cite{BaBar:2009hkt}, LEP \cite{Hook:2010tw}, and E137 \cite{Bjorken:1988as} -- the exclusion regions for BABAR and E137 are based on the reinterpretation in \cite{Izaguirre:2017bqb}. The dashed curves stand for projected limits from BDX \cite{BDX:2016akw}, LDMX \cite{LDMX:2018cma}, MiniBooNE \cite{MiniBooNEDM:2018cxm} and Belle II \cite{Belle:2021ysv}, also computed in Ref. \cite{Izaguirre:2017bqb}.
    }
    \label{fig:reach}
\end{figure*}

The signal topology from \Eq{eq;signal-topology} consists of one primary muon track, and two additional charged tracks originating from the same displaced vertex.
We also require that all three tracks in the event pass through all three tracking layers in their module of origin to ensure high event reconstruction efficiency.
Two of these charged tracks must be reconstructed from a common displaced vertex between the target and the first tracking layer.
These displaced tracks must also have an opening angle larger than $1~\rm mrad$ to  successfully reconstruct the displaced vertex \cite{Umberto:2021private}. 
As we will see in the next section, the two tracks from the displaced vertex must be either an $e^+ e^-$ or $\mu^+ \mu^-$ pair in order to reject SM backgrounds with the PID system.
The primary muon must also be identified by the ECAL and the Muon filter, as in the original proposal of the MUonE experiment.
Given the expected $z$-direction spatial resolution $\delta z \simeq 1~\rm mm$~\cite{Umberto:2021private,Galon:2022xcl}, we further require that this displaced vertex is at least $10 \delta z$ away from both the target and the first tracking layer.
Thus, we require that the longitudinal endpoints of the fiducial $\chi_2$ decay region satisfy
\begin{align}
\label{eq:deltaz}
    25~\mathrm{mm} \leq z_{\chi_2~\text{decay}} \leq 140~\mathrm{mm},
\end{align}
as measured from the interaction point within a given target module.
In Fig.~\ref{fig:close}, we show a schematic overhead view of a signal event whose final state tracks satisfy our geometric selection criteria.

Beyond geometry, we must also ensure that these tracks correspond precisely to the particles from our signal prediction (one beam muon and two displaced dileptons), as opposed to fakes from other particles. Furthermore, we must also ensure that no additional neutral hadrons are produced in association with three charged tracks, as our scenario predicts no additional hadronic activity in the final state. 
For these purposes, the ECAL and Muon system at the end of the beam line are essential for vetoing backgrounds. For example, if the displaced tracks arise from charged pions instead of electrons, the energy deposited in the ECAL can discriminate between these cases and the absence of muons in the Muon system can rule out the possibility that the displaced tracks arise from dileptons. Alternatively, even if a given event has exactly one beam muon and a displaced dilepton pair, if there are additional energy deposits in the ECAL that do not correspond to any charged tracks (e.g. from a neutron), it can also be vetoed as background.

However, as particles lose energy traversing the tracking layers, the ECAL PID capability can deteriorate, thereby reducing its ability to veto backgrounds. Muon and electron can scatter off the material when penetrating the targets and the tracking layers in each module before they arrive the ECAL.
To mimic the detector response of the tracking layers and the ECAL, as discussed in Ref.~\cite{Galon:2022xcl} based on details of energy loss~\cite{walters2017stopping,Workman:2022ynf}, we consider a conservative\footnote{This requirement was based on internal communication~\cite{Umberto:2021private} and was applied as the most conservative choice in one of the previous study~\cite{Galon:2022xcl}} energy threshold of $E > 5~\rm GeV$ for each of the three charged tracks.
For the same reason, only the last 5 modules are considered in this work to ensure sufficiently high ECAL PID efficiency; events originating in modules further upstream risk significant energy loss from the other 35 target and tracking modules.
Furthermore, since the ECAL has a large surface area, charged tracks with energy above the energy threshold are guaranteed to enter the ECAL if they pass through the three tracking stations for these last 5 modules.

Our full selection criteria are described in Table~\ref{tab:cuts} and these requirements ensure that 
our search is free from SM backgrounds. In the Supplemental Material, we provide a detailed description of our background simulation to justify these numbers.

{\bf Signal Simulation.}
To compute the number of expected signal events in MUonE, we use \verb|FeynRules| \cite{Alloul:2013bka} to construct the inelastic dark matter model.
We then use \verb|CalcHEP 3.8.9| \cite{Belyaev:2012qa} to generate Monte Carlo (MC) events for $(\mu^-\, \text{Be} \to \mu^-\, \text{Be}\, A')$ with a muon beam energy of 160\,GeV and kinematic distributions for the $\chi_2 \to \chi_1 e^+ e^-$ and $\chi_2 \to \chi_1 \mu^+ \mu^-$ decay channels. For each point, we generate $5 \times 10^4$ signal events. 
The production events are then re-weighted according to the nuclear form factor of beryllium (an analytic formulation of this process is described in \cite{Chen:2017awl}).
These events are then filtered through the series of cuts discussed in the previous section and listed in Table~\ref{tab:cuts}.

In Fig.~\ref{fig:aprime-dist} we show distributions of $\Ap$ production
in our signal simulation. The left panel shows the total cross section for $\mu N\to \mu N \Ap$ production normalized to $\epsilon = 1$ and the right panel shows the energy distribution of recoiling muons following $\Ap$ production for various values of $m_{\Ap}$. In Fig.~\ref{fig:x2-dist} we also show energy angle distribution histograms for the simulated $\chi_2$ produced in $\Ap \to \chi_1 \chi_2$ decays at MUonE.  We note that our Be form factor becomes uncertain for $t > 4m_p^2$, where $t$ is the momentum transfer squared \cite{Bjorken:2009mm}. We have verified that our results are unchanged when we apply a cut to remove events that violate this inequality.

{\bf Results and Discussion.}
In Fig.~\ref{fig:reach} we present our main results, which demonstrate that, for a design $\mu - \rm Be$ luminosity of $4.7 \times 10^3$ pb$^{-1}$\footnote{The total $\mu -\rm Be$ luminosity is smaller than the luminosity for $\mu-e$ scattering by a factor of $1/4$. We further divide this number by 8 since we only use the last 5 (out of 40) modules.}, MUonE (solid red curve) can probe much of the parameter space compatible with thermal freeze out (solid black curve) for a variety of different mass splittings.
Our projection is defined with a 95\% confidence interval using the
event selection criteria from Table \ref{tab:cuts}, which corresponds to 3 signal events that pass the cuts. For these results, we assume that SM backgrounds are negligible, which is justified by our background simulation described above.
Also shown are existing limits from BaBar and E137 computed in Ref. \cite{Izaguirre:2017bqb}.
Note that, for smaller values of $\Delta$ than those shown here, the $\chi_2$ decay length is much longer than the fiducial decay region, so the MUonE would only be sensitive to very large couplings that compensate for this decrease in $\Delta$; for $\Delta \lesssim 0.2 m_1$ the necessary values of $y$ needed to compensate for this suppression are all excluded.
Thus, in Fig.~\ref{fig:reach}, we only show splittings in the $\Delta = (0.2-0.4) m_1$ range.
% This is explained by the long decay length under the small splitting, which could be only compensated by a larger coupling.
% For $\Delta < 0.1 m_1$, MUonE is not expected to provide any good sensitivity.

The shaded gray regions in Fig.~\ref{fig:reach} represent existing limits on 
this scenario from the BaBar monophoton search \cite{BaBar:2009hkt,Izaguirre:2017bqb}, the E137 electron beam dump search for long-lived particles \cite{Bjorken:1988as,Izaguirre:2017bqb}, and LEP precision electroweak constraints on kinetically mixed dark photons \cite{Hook:2010tw}.
The dashed curves represent future projections for BDX \cite{Izaguirre:2013uxa,BDX:2016akw}], LDMX \cite{Izaguirre:2014bca,Berlin:2018bsc,LDMX:2018cma}, MiniBooNE \cite{MiniBooNEDM:2018cxm,Izaguirre:2017bqb}, and Belle II \cite{Belle-II:2018jsg,Izaguirre:2017bqb}

In summary, we have shown that the MUonE experiment can powerfully probe thermal relic dark matter in models with inelastic mass splittings. In such models, dark states are produced in muon-beryllium scattering and the heavier state decays semi-visibly to yield a displaced dilepton pair downstream of the target. Our approach is fully parasitic with the proposed MUonE experimental setup and requires no additional equipment. 
Furthermore, our results demonstrate the importance of the MUonE ECAL system, which is currently at risk of being eliminated from the experimental design \cite{Ulrich:2024private}. In our search, the ECAL plays an essential role in rejecting SM backgrounds for the dark matter signal and expanding the MUonE experimental program. Thus, we strongly encourage the collaboration to keep this component as part of the full setup.

{\bf Acknowledgments.} 
We thank David Shih and Yannick Ulrich for helpful conversations. We also thank John Beacom for feedback on the manuscript. Fermilab is operated by Fermi Research Alliance, LLC under Contract No.~DE-AC02-07CH11359 with the U.S. Department of Energy, Office of Science, Office of High Energy Physics.
I.R.W. is supported by DOE distinguished scientist fellowship grant FNAL 22-33.
I.R.W. is grateful to the computing cluster support from the Rutgers CMS group. D.R. is supported by the U.S. Department of Energy, Office of Science, Office of Workforce Development for Teachers and Scientists, Office of Science Graduate Student Research (SCGSR) program. The SCGSR program is administered by the Oak Ridge Institute for Science and Education for the DOE under contract number DE‐SC0014664.

\bibliographystyle{utphys3}
\bibliography{biblio}

% Switch to single-column format for supplemental material
\onecolumngrid

\appendix
\section*{Supplemental Material}
\renewcommand{\theequation}{S\arabic{equation}}
\setcounter{equation}{0}

\noindent In this supplement we identify and simulate the backgrounds for our proposed search strategy described above. 
Standard Model processes that fake our signal fall into two distinct  categories:
\begin{itemize}
    \item {\bf Mis-reconstructed events: } backgrounds can arise from SM events with charged tracks that originate within the target, but are mis-reconstructed as displaced vertices. This category also includes events in which SM particles produced inside the target interact later to generate additional $e^+e^-$ tracks within the downstream tracking layers.
    By requiring the decay volume to have a $10~\delta z$ displacement away from both the target and the first tracking layer, this category of events can be safely rejected~\cite{Galon:2022xcl,Umberto:2021private}.

    \item{\bf Long-lived hadrons:}
     genuine displaced vertex can arise from long-lived hadrons that decay to yield dileptons in between the target and first tracking layer. These particles are produced in inelastic muon-nucleon scattering and decay downstream of the target to yield high multiplicities of charged tracks (more than two charged tracks plus the beam muon), which can easily be vetoed.

\end{itemize}

To estimate the number of background events from the second category, we run \verb|Pythia 8.307|~\cite{Bierlich:2022pfr} to simulate muon-nucleon scattering at MUonE.
The muon-nucleon scattering can be categorized by $Q^2$, the momentum transfer to the nucleus.
A large $Q^2$ corresponds to deep-inelastic scattering (DIS), which can be modeled using perturbative QCD, and
in this regime, the muon deflection angle has a minimum value.
Below this threshold, soft QCD production dominates, in which case the muon deflection angle is bounded from above.

We generate $1.5\times 10^{10}$ soft QCD events and $6 \times 10^8$ deep-inelastic scattering (DIS) events; both samples greatly exceed the expected number of such events in each category for the MUonE luminosity.\footnote{For muon-Be scattering, the inclusive soft QCD cross section is $\sigma \approx 8.2 \times 10^5~\rm pb$, while the inclusive DIS cross section is $\sigma \approx 3.4 \times 10^4~\rm pb$, where the separation between these regimes is defined at the threshold momentum transfer $Q^2= (2~\rm GeV)^2$ } 
The minimum simulated $Q^2$ delivered to the target for DIS (a free parameter in \verb|Pythia|) is chosen so that the differential cross section for hadroproduction is continuous across the soft-QCD and DIS kinematic regimes. 
Using these samples, we veto any simulated SM events that contain tracks beyond the three that can fake our signal topology (one track from the beam muon and two displaced tracks from the dilepton pair).
Such additional tracks can arise from charged particles with energies above the 5 GeV lepton energy threshold (see Table \ref{tab:cuts}).
In addition, any neutral particles above this energy threshold (except neutrinos) can be detected and vetoed using the downstream ECAL.
Importantly, we find that this conclusion is insensitive to the choice of energy threshold, varying from $10~\rm MeV$ to $5~\rm GeV$.
Note that the vast majority of the background events from hadronic processes arise from $K^0 \to \pi^+ \pi^-$ or $\Lambda \to p \pi^-$ decays.
In these cases, none of the charged tracks faking the displaced vertex are real electrons or muons.
Our simulation finds approximately 3500 such background events are expected to fake the signal event topology before running the PID system; these can be safely rejected as long as the fake rate of the PID is lower than $\sim 2\%$ per particle species.

There are potentially important backgrounds from a variety of possible sources: 

\begin{itemize}
    \item {\bf Real lepton + fake hadron:} processes involving one real electron or muon and a second charged particle produced together from a common displaced vertex  (e.g. $K^0 \to \pi^+ e^- \bar{\nu}_e$ or $\Lambda \to p e^- \bar{\nu}_e$) can fake our signal if the charged hadron is misidentified as an electron or muon.
Using the same simulation details described above, we find that MUonE can expect approximately $\sim$ 6 SM events for $\sim 10^{16}$ muons on target, and these can be safely rejected using PID as long as the per-particle fake rate is lower than $18\%$.

\item{\bf Real dileptons:}  there are also SM processes that result in displaced $e^+e^-$ pairs with displaced vertices (e.g. $K^0 \to \pi^0 \pi^0$ with one $\pi^0 \to e^+ e^- \gamma$ decay), which also fakes our signal topology.
However, such production processes are always accompanied by high-energy photons (coming from the $\pi^0 \to e^+ e^- \gamma$ decay and from the other $\pi^0$ decay in the event), so these processes can be efficiently vetoed; our simulation shows that no such background events passed our selection criteria.
This conclusion is consistent with the results of Ref.~\cite{Galon:2022xcl}, which found similarly low SM backgrounds for MUonE sensitivity to new forces.

\item 
{\bf QED hadroproduction:} finally, electromagnetic muon-nucleus scattering can occasionally produce a $K^0$ or $\Lambda$ to fake our signal via $\mu N \to \mu N \gamma^* (\gamma^* \to \bar{q}q)$. 
This process, however, is suppressed by both the production cross section and the branching ratio for the $\bar{q}q$ to combine into a $K^0$ or a $\Lambda$.
We estimate that such processes contribute less than 1 event per $10^{16}$ muons on target.
\end{itemize}

\noindent Based on these considerations,  we conclude that SM backgrounds are negligible for our selection criteria.

\end{document}